\documentclass[a4paper,fleqn,usenatbib]{mnras}

\usepackage{newtxtext}
\usepackage{newtxmath}
\usepackage{natbib}
\usepackage[T1]{fontenc}
\usepackage{aecompl}

\usepackage{graphicx}							
\usepackage{bm}									
\usepackage{amsmath}							
\usepackage{amssymb}							
\usepackage{physics}							
\usepackage[nameinlink, capitalize]{cleveref}	
\usepackage{array}
\graphicspath{{graphics/}}

\newcommand{\erad}{\theta_\mathrm{E}}						
\newcommand{\reff}{r_\mathrm{eff}}							
\newcommand{\imra}{{\theta_\mathrm{r}}}						
\newcommand{\flra}{{f_\mathrm{r}}}							
\newcommand{\srcf}{{f_\mathrm{s}}}							
\newcommand{\posa}{\phi_\mathrm{L}}							
\newcommand{\etheta}{\theta_\varepsilon}					
\newcommand{\sg}{\sigma_\gamma}								
\newcommand{\nim}{N_\mathrm{im}}							
\newcommand{\eu}{\mathrm{e}\mkern1mu}						
\newcommand{\iu}{\mathrm{i}\mkern1mu}						

\title[Mass profiles from lensing II]{Galaxy mass profiles from strong lensing II: The elliptical power-law model}

\author[C. M. O'Riordan et al.]{
	C. M. O'Riordan,$^{1}$\thanks{E-mail: conor.oriordan15@imperial.ac.uk}
	S. J. Warren,$^{1}$
	D. J. Mortlock$^{1,2,3}$
	\\
	$^{1}$Astrophysics Group, Blackett Laboratory, Imperial College London, London, SW7 2AZ, United Kingdom\\
	$^{2}$Department of Mathematics, Imperial College London, London, SW7 2AZ, UK\\
	$^{3}$Department of Astronomy, Stockholm University, Albanova, SE-10691 Stockholm, Sweden
}
\date{Accepted XXX. Received YYY; in original form ZZZ}
\pubyear{2020}

\begin{document}
\label{firstpage}
\pagerange{\pageref{firstpage}--\pageref{lastpage}}
\maketitle

\begin{abstract}
\noindent	
We present a systematic analysis of the constraints $\sg$ on the mass profile slope $\gamma$ obtainable when fitting a singular power-law ellipsoid model to a typical strong lensing observation of an extended source. These results extend our previous analysis of circular systems, Paper I. We draw our results from 676 mock observations covering a range of image configurations, each created with a fixed signal to noise ratio $S=100$ in the images. We analyse the results using a combination of theory and a simplified model which identifies the contribution to the constraints of the individual fluxes and positions in each of the lensed images. The main results are: 1. For any lens ellipticity, the constraints $\sg$ for two image systems are well described by the results of Paper I, transformed to elliptical coordinates; 2. We derive an analytical expression for $\sg$ for systems with the source aligned with the axis of the lens; 3. For both two-image systems and aligned systems, $\sg$ is limited by the flux uncertainties; 4. The constraints for off-axis four-image systems are a factor of two to eight better, depending on source size, than for two-image systems, and improve with increasing lens ellipticity. We show that the constraints on $\gamma$ in these systems derive from the complementary positional information of the images alone, without using flux. The complementarity improves as the offset of the source from the axis increases, such that the best constraints $\sg<0.01$, for $S=100$, occur when the source approaches the caustic.
\end{abstract}


\defcitealias{ORiordan2019}{Paper I}
\defcitealias{Tessore2015a}{TM15}


\section{Introduction}
\label{sec:introduction}

The measurement of the mass profile of a galaxy acting as a
gravitational lens is important in at least four areas of astronomy: i) the
determination of the Hubble constant using the time-delay method
\citep[e.g.][]{Suyu2010}; ii) the detection of mass substructure in galaxy
haloes \citep[e.g.][]{Vegetti2014}; iii) the accurate measurement of the
mass-to-light ratio in galaxies, which may be used to estimate the
stellar initial mass function \citep[e.g.][]{Auger2010}; iv) testing theories of
the formation and evolution of galaxies \citep[e.g.][]{Koopmans2009}. 

The prevailing view is that strong lensing (i.e. when multiple images are formed) on its own yields only limited information on the mass
profile, and that supplementary information from dynamics is required.
The goal of this series of papers is to reexamine this question. 
The first paper in the series, \citet[][hereafter
`\citetalias{ORiordan2019}']{ORiordan2019} includes a literature review of the
topic.  As noted there, much of the theory of how strong lensing can
constrain the mass profile makes reference only to measurement of the
positions of the images, and the theory of the constraints provided by
flux information is underdeveloped (although see below). For point sources, i.e. quasars, the flux information cannot be used because it may be
affected by variability or microlensing, but for extended sources the
flux information provides additional constraints. In \citetalias{ORiordan2019} we
examined the case of a circular lens with a singular power-law mass
profile, $\rho\propto r^{-\gamma}$. In the current paper we extend this analysis to the
elliptical case.

This series of papers extends the theory of measurement of the lens mass profile in fitting a parametric global model for the mass distribution. There has been substantial progress recently in the development of an alternative approach, termed model-independent, which is based on the insight that the observables constrain only certain local properties of the potential \citep{WagandBart,Wagner1,Tessore2017, Wagner2, Wagnerrev}. The observables include positions and fluxes, as well as ellipticities, orientations and time delays. The properties constrained are derivatives and ratios of derivatives of the potential. These two approaches are complementary and it is a requirement of any global fit to satisfy all the model-independent constraints.

We now briefly summarise the contents of \citetalias{ORiordan2019}. For a circular lens modeled with a power-law mass profile there are four parameters to be determined: the mass profile slope $\gamma$; the Einstein angle $\erad$;  the source angular position $\beta$; and the source flux $\srcf$. Since it is the mass profile we are interested in, the question is how accurately can the slope be measured, i.e. what is $\sg$ for different image configurations.
The images provide two angular positions $\theta_1$ (inner) and $\theta_2$ (outer), and two fluxes $f_1$ and $f_2$, so there are potentially sufficient measurements to measure the four parameters. Working with ratios $\imra=\theta_1/\theta_2$ and $\flra=f_1/f_2$ eliminates $\erad$ and $\srcf$. We then showed that $\imra$ alone, or $\flra$ alone, provides constraints on a combination of the mass profile slope $\gamma$ and the source position $\beta$ (\citetalias{ORiordan2019}, Fig. 3). These constraints are complementary, so by combining the position and flux information both parameters can be measured. 

We found that the uncertainty on the measured value of $\gamma$ is dominated by the uncertainty on $\flra$, and that for this case the positions may be treated as measured exactly provided the source size is much smaller than $\erad$. With this assumption we were able to derive the uncertainty on $\gamma$ as a function of: $\gamma$ itself; the position ratio $\imra$; and the summed signal-to-noise ratio (S/N) in the two images, parameterised by $S$. For the isothermal case $\gamma=2$ this relation reduces to the simple expression:
\begin{equation}
\label{eq:sigmagamma}
	\sg=\frac{2\sqrt{\imra}}{S(1-\imra)}.
\end{equation}
For $S=100$, which is quite commonly achieved in HST images, and for $\imra=0.5$, this formula implies $\sg=0.03$, i.e., the profile slope may be measured very accurately.

We confirmed these results using a series of mock observations. We modeled the source as circular, with a S\'{e}rsic surface-brightness profile, imaged by a circular power-law lens. Two sets of mock observations were created with different source sizes. Noise was added as appropriate, to produce images with $S=100$. We fitted an elliptical lens, i.e. the singular power-law ellipsoid (SPLE), and an elliptical source, and the uncertainty $\sg$ was recorded. The lens position is assumed known, i.e. in a real image it would be measured as the centroid of the light of the lens galaxy.\footnote{We follow this assumption in all this work. In principle the lens centre could be added as two extra parameters $x_l, y_l$, which would change the results, but this possibility is not considered here.}  The fitted model has a total of 11 parameters, seven for the source and four for the lens (details are provided in \cref{sec:method}). Despite the larger number of parameters, we found very close agreement between the value of $\sg$ obtained by fitting  the mock observations and the prediction of the simplified theoretical analysis. This demonstrates that the theory captures all the relevant features of the problem, and that the results are independent of the structure of the source. 
 
In the current paper we extend the analysis to the elliptical case. As for the circular case we present a systematic treatment of the available constraints on the mass profile slope, using lensing data alone, i.e., without dynamics. 

The lensing features of singular power-law ellipsoid (SPLE) mass distributions were first found analytically by \citet{Bourassa1975}, to which \citet{Bray1984} added a minor correction. \citet{Kormann1994} found analytic expressions for the lensing properties for the particular SPLE case $\gamma=2$ (i.e. the isothermal model). Later \citet{Schramm1990} and \citet{Barkana1998} devised numerical routines that allow the analysis of the general SPLE case. More recently, \citet{Tessore2015a} found elegant analytic expressions for the general case, which we will make use of in this work. The isothermal SPLE is the choice of mass model in many galaxy-galaxy lensing observational studies including in SLACS \citep{Bolton2008a} and in the BELLS GALLERY survey \citep{Shu2016}.

For the circular power-law case we presented a complete analytical treatment of the constraints on $\gamma$, but for the SPLE this is possible for only one particular configuration. Therefore for a complete treatment we must rely on analysing mock observations. In this paper we present a set of mock observations spanning a range of source positions and a range of lens ellipticities. In each we fit the same $11$ parameter model as before and find the $1\sigma$ uncertainty on the slope. These results reveal that the relation between $\sg$ and the image configuration is more complex for the elliptical case. 
We identify three separate regimes, listed in \cref{table:regimes}, and find that the origin of the constraint on the slope in relation to the observables is different in each regime. The observables contributing to the constraint $\sg$ in each regime are listed in the table. Regime 1 is the case of two images, where the source is located outside the astroid caustic. Regime 2 is where the source lies on the optical axis, producing a four-image Einstein cross configuration. We refer to these as `aligned systems'. This regime is tractable analytically. Regime 3 is the general case of four images, where the source lies off axis but inside the astroid caustic. To gain a detailed understanding of how a particular configuration constrains $\gamma$ we make use of a simplified modelling apparatus which we call the position/flux model. With this we fit directly to the positions and fluxes rather than the image pixel values. The power of this approach is that we can determine exactly which of the observables are responsible for the constraints in the different regimes, because any observables \---\ positions or fluxes \---\ can be enabled or disabled at will in the fitting procedure. 

\begin{table}
\centering
	\begin{tabular}{cccc}
\textbf{Regime}	&	\textbf{Images} & \textbf{Configuration} & \textbf{Observables}\\
		\hline
 1 & 2 & all          & fluxes+positions \\
 2 & 4 & aligned  & fluxes+positions \\
 3 & 4 & off-axis  & positions \\ \hline
 	\end{tabular}
	\caption{\label{table:regimes}The three regimes into which the different image configurations are divided. The manner in which the observables lead to constraints on $\gamma$ is different in each regime.}
\end{table}
 
The paper is organised as follows. \Cref {sec:theory} details the relevant properties of the SPLE. \Cref {sec:method} gives a brief summary of the method for creating and constraining a mock observation and \cref {sec:results} gives the results for the mock observations. In \cref {sec:aligned-systems} we derive an expression for the uncertainty in the aligned four-image systems. In \cref {sec:point-source-model} we use the position/flux model to treat the general, off-axis, four-image case as well as the two-image case.  In \cref {sec:conclusions} we present a discussion of the results and a summary.

\section{The singular power-law ellipsoid}
\label{sec:theory}
The dimensionless mass density profile, or convergence, for the SPLE is given by
\begin{equation}
	\label{eq:convergence}
	\kappa(\etheta) = \frac{3-\gamma}{2}\left(\frac{b}{\etheta}\right)^{\gamma-1},
\end{equation}
where $\gamma$ is the exponent of the power-law radial variation of density in three dimensions, $\rho\propto r^{-\gamma}$.
This equation is identical in form to the circular power-law case we analysed in \citetalias{ORiordan2019} except the coordinate used is now an elliptical radius $\etheta$, defined
\begin{equation}
	\label{eq:elliptical-radius}
	\etheta^2 = q^2\theta_x^2+\theta_y^2,
\end{equation}
where $q$ is the ratio of the minor to major axis of the isodensity contours, which are homoeoidal ellipses. The ellipticity is defined $\varepsilon=1-q$. The Einstein radius has been replaced with the more generic lensing strength, $b$, defined such that the total projected mass $M$ within an angular radius $\etheta=b$ is
\begin{equation}
	M(b) = \Sigma_\mathrm{crit}\pi b^2 D_\mathrm{d}^2/q,
\end{equation} 
where $D_\mathrm{d}$ is the angular diameter distance to the lens and $\Sigma_\mathrm{crit}$ is the critical density, given by
\begin{equation}
	\Sigma_\mathrm{crit}=\frac{c^2}{4\pi}\frac{D_\mathrm{s}}{D_\mathrm{ds} D_\mathrm{d}}
\end{equation}
where $D_\mathrm{s}$ and $D_\mathrm{ds}$ are the angular diameter distances to the source and from the lens to the source respectively \citep[e.g.][]{Schneider1992}.
With this definition a lens with lensing strength $b$ has the equivalent Einstein radius $\erad=b/\sqrt{q}$.

\begin{figure}
	\includegraphics[width=1.0\columnwidth]{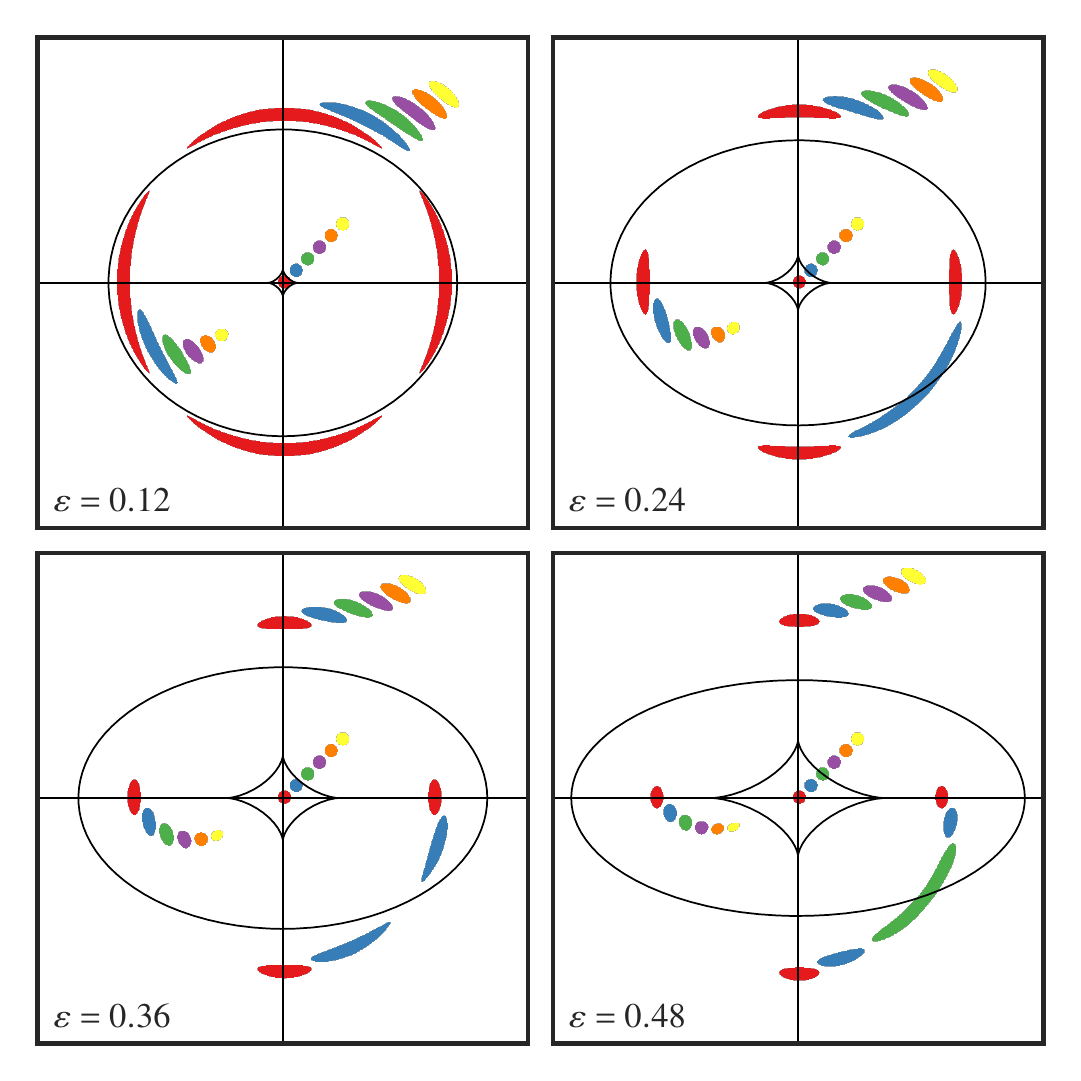}
	\caption{\label{fig:sple-diagram} The image structure for an isothermal ($\gamma=2$) SPLE at four different ellipticities. The set of discs in the centre of each frame is a set of six sources covering $0\leq\beta\leq0.5$. Each source's images are plotted in the same colour.}
\end{figure}

We restate here the main results of \citet{Tessore2015a} as they will be useful later in the paper and we follow their complex formalism where $z=\theta_x+\iu \theta_y$. The deflection angle $\alpha=\alpha_x+\iu \alpha_y$ for the SPLE is given by
\begin{equation}
	\label{eq:deflection-angle}
	\alpha(\etheta,\varphi)=\frac{2b}{1+q}\left(\frac{b}{\etheta}\right)^{\gamma-2}\eu^{\iu\varphi}{}_2F_1\left(1,{\textstyle\frac{\gamma-1}{2}};{\textstyle\frac{5-\gamma}{2}},{\textstyle-\frac{1-q}{1+q}\eu^{\iu 2\varphi}}\right),
\end{equation}
where $\varphi=\mathrm{arctan}(q\theta_x,\theta_y)$ and ${}_2F_1$ is the Gaussian hypergeometric function. By calculating the shear, one obtains the (inverse) magnification as a function of convergence and deflection angle
\begin{equation}
	\label{eq:magnification}
	\mu^{-1}=1-2\kappa(z)\left[1-(2-\gamma)\frac{\theta_x \alpha_x + \theta_y \alpha_y}{\theta^2}\right]-(2-\gamma)^2\frac{\abs{\alpha}^2}{\theta^2},
\end{equation}
where $\theta^2=\theta_x^2+\theta_y^2$ is the radius in circular coordinates at $z$. For an isothermal slope, where $\gamma=2$, we get $\mu^{-1}=0$ when $\kappa=1/2$. From \cref {eq:convergence} we then see that $\mu\rightarrow\infty$ as $\etheta\rightarrow b$. The ellipse where $\etheta=b$ is therefore the critical curve for the isothermal SPLE.

With $z'=\beta_x+\iu\beta_y$ as the complex plane source coordinate, the lens equation is
\begin{equation}
	\label{eq:lens-equation}
	z'=z-\alpha(z).
\end{equation}
With this we can transform the critical curve into the source plane and obtain the caustic. The caustic is the curve in the source plane on which sources are infinitely magnified and its exact form can be computed numerically from \cref {eq:deflection-angle}. In the isothermal SPLE the caustic is astroid shaped and increases in size with ellipticity, vanishing to a point at the origin for $q\rightarrow1$. Sources outside the caustic are lensed into two images, one inside and one outside the critical curve. Sources inside the caustic form four images, two inside and two outside the critical curve. \cref {fig:sple-diagram} illustrates the image structure for an isothermal SPLE at different ellipticities.

\begin{figure*}
	\includegraphics[width=1.0\textwidth]{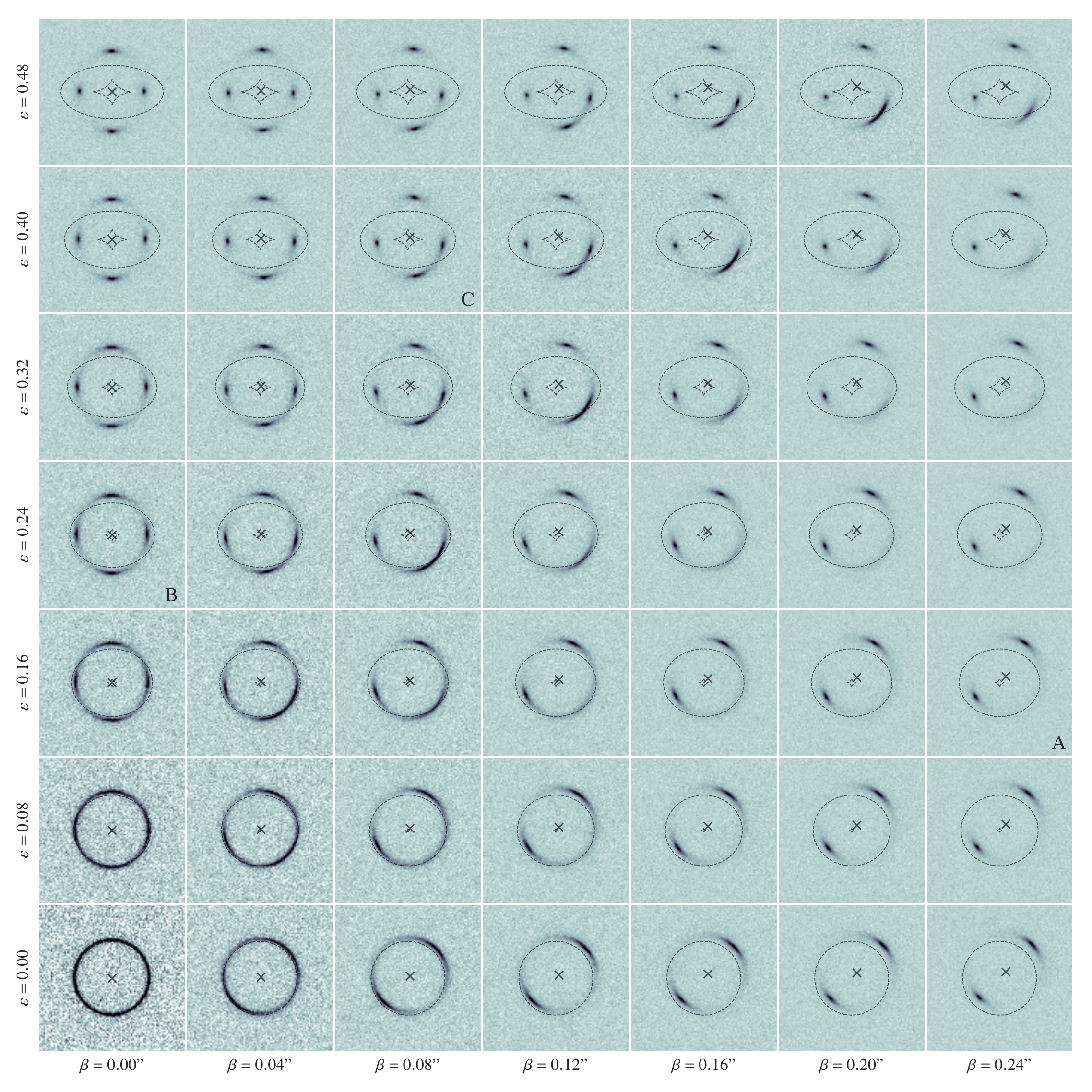}
	\caption{\label{fig:lens-grid} A small sample of the 676 mock observations used in this  paper. Caustics and critical lines are shown as dashed curves on each image plane. To more clearly show the behaviour in the four image systems, only a truncated range of source positions $(0$ arcsec $<\beta<0.24$ arcsec$)$ is plotted. The source position is marked with a cross. The image planes are $6$ arcsec $\times$ $6$ arcsec in modelling, but for clarity only the inner $4$ arcsec $\times$ $4$ arcsec region is shown here. The systems labelled A, B and C are examples of each regime that we identify and the posterior density distributions for these systems are in \cref{fig:corner-plots}.}
\end{figure*}

\begin{figure*}
	\includegraphics[width=1.0\textwidth]{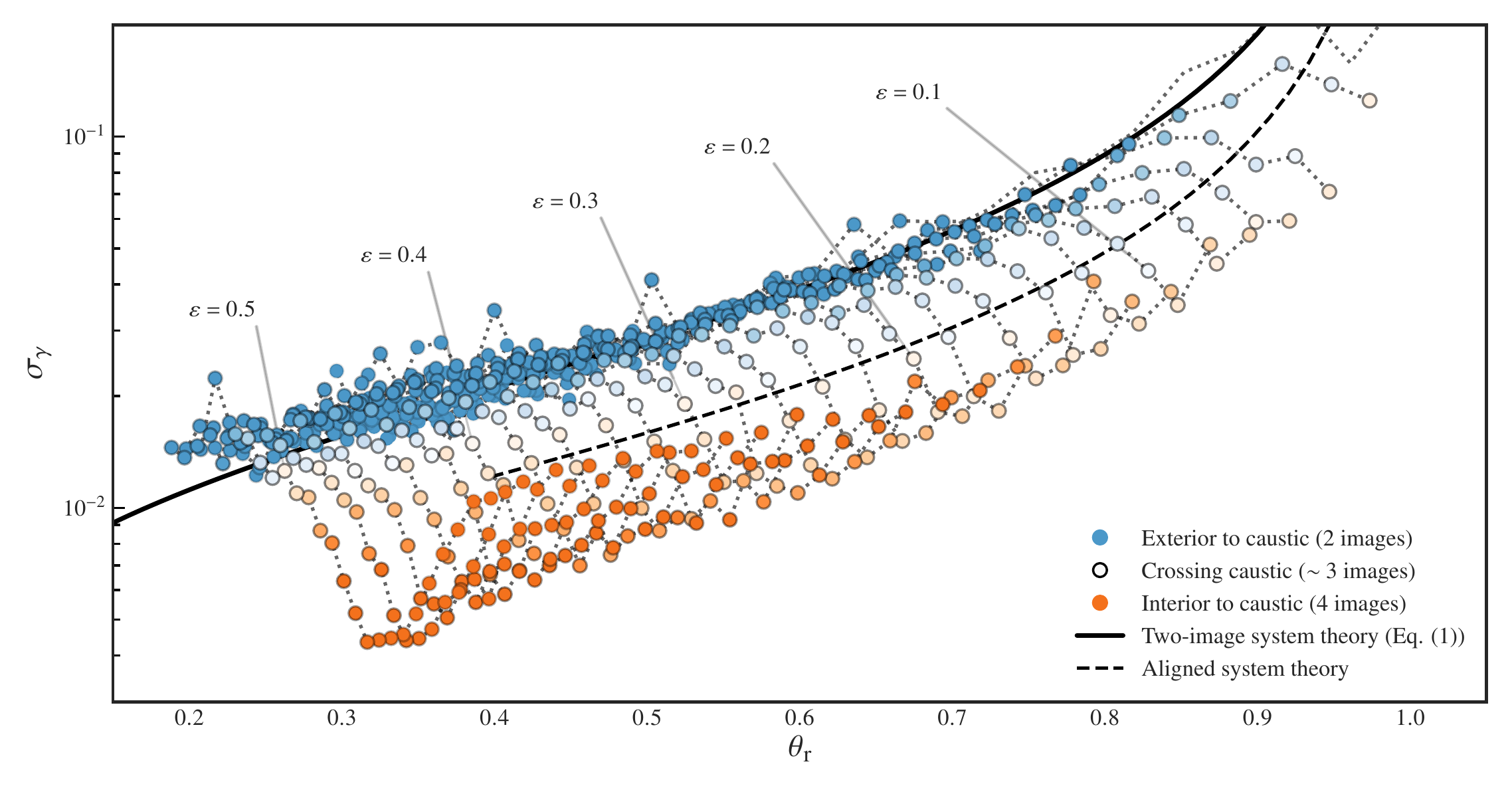}
	\caption{\label{fig:full-results} The uncertainty on the mass-profile slope $\sg$ for all $676$ mock observations as a function of image position ratio $\imra$. Systems are coloured according to their source's position in relation to the caustic. Systems with the same ellipticity are connected by dotted lines. Along these tracks of ellipticity, systems with $\beta=0$ arcsec are furthest right and those with $\beta=0.5$ arcsec are furthest left. The ellipticity of every fifth track is labelled. The solid black line is the expected uncertainty for a circular system with image position ratio $\imra$ and S/N $=100$, \cref {eq:sigmagamma}. The dashed curve is the result of the analysis for aligned systems in \cref{sec:aligned-systems}.}
\end{figure*}

\section{Method}
\label{sec:method}
The method for creating a mock observation and constraining its parameters is detailed fully in Section 3 of \citetalias{ORiordan2019}. The process is identical in this paper but we briefly summarise the important features here.

The lens is modelled as an isolated, transparent SPLE with four parameters $(b, \gamma, q, \posa)$, where $\posa$ defines the orientation of the major axis of the lens. The source surface brightness is modeled by a S\'ersic profile with seven parameters $(\beta_x,\beta_y,\reff,I_0,n,q_\mathrm{s},\phi_\mathrm{s})$.  The lenses have $b=\sqrt{q} \arcsec, \gamma=2, \posa=0$, and sources are created with $\reff=0.1\arcsec, I_0=1, n=2, q_s=1$ (i.e. circular). After creating a model image plane we add Gaussian noise to each pixel such that $S=100$ integrated within a well-defined mask containing most of the signal. By fixing $S$ within the mask, `all lenses are created equal', so that it is possible to quantify the relative effectiveness of different configurations for constraining $\gamma$. The model parameters are then constrained via ensemble Markov chain Monte Carlo (MCMC) sampling. The uncertainty on the profile slope, $\sg$, is taken to be the mean distance from the 16th and 84th percentiles to the median in the posterior samples. For a normal posterior this is equivalent to the $1\sigma$ uncertainty on $\gamma$.

For a real galaxy treating the lens as isolated will give incorrect results if there are significant contributions to the potential from other galaxies nearby or along the line of sight. Perturbations from other galaxies need to be treated explicitly unless their effect can be shown to be negligible in the context of the scientific question being addressed.  In this regard we consider that modelling a perturbing galaxy by adding external shear only, without convergence, is incorrect, since the external convergence will affect the parametric fit. This point is demonstrated explicitly with simulations by \citet{McCully2017}, who show how treating the shear and convergence from perturbing galaxies in a self-consistent manner removes biases in the fitted parameters of the lens.

\section{Results from Mock Observations}
\label{sec:results}

Our results are drawn from a set of $676$ mock observations. A small sample of images is shown in \cref {fig:lens-grid}. The full set spans a uniformly spaced grid of $26$ source positions $(0$ arcsec $\leq\beta\leq0.5$ arcsec$)$ and $26$ ellipticities $(0.0\leq\varepsilon\leq0.5)$. Here $\beta^2=\beta_x^2+\beta_y^2$, and the source is moved diagonally along the line $\beta_x=\beta_y$ (as in \cref {fig:sple-diagram}) such that a system with source position $\beta$ has $\beta_x=\beta_y=\beta/\sqrt{2}$. This choice of source positions and ellipticities means that the majority of strong lens systems with a single source will resemble a system somewhere on this grid. This set of configurations does not include those where the source lies in the cusp of the caustic. This is a special case and so we exclude it here for the sake of brevity. However, in analysing a number of cusp systems we have found them to fit with the more general theories established in the next two sections.

The measurements of $\sg$ for all 676 mock observations are summarised in \cref {fig:full-results}, plotting $\sg$ as a function of $\imra$. For a general lens system with multiple images we calculate the elliptical radius, according to \cref {eq:elliptical-radius}, at each image. We then define $\imra$ as the ratio of the minimum over the maximum elliptical radius. The figure is colour coded by image multiplicity, with blue representing 2-image systems, orange representing 4-image systems, and a gradation in colour saturation as the source crosses the astroid caustic and the multiplicity changes between two and four. It is immediately apparent that even though all systems are created with the same total S/N, $S=100$, the constraints on $\gamma$ are better for 4-image systems than for 2-image systems, by up to a factor of 5.

\begin{figure}
	\includegraphics[width=1.0\columnwidth]{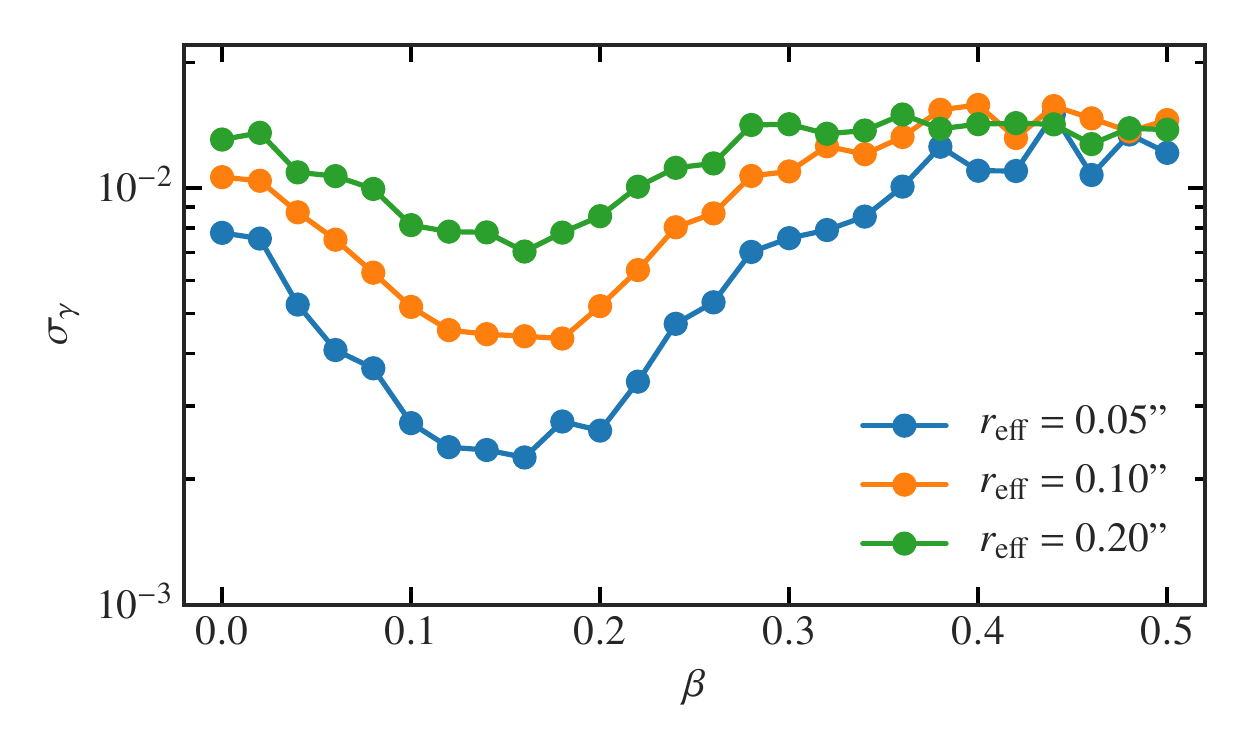}
	\caption{\label{fig:source-size} The uncertainty on the slope measured from the mock observations as a function of source position, for three different source sizes, for lenses of $\varepsilon=0.5$. The curves for other values of $\varepsilon$ show similar behaviour.}
\end{figure}

The majority ($73\%$) of the systems in the population have two images, and belong to Regime 1. The black solid line is the relation \cref {eq:sigmagamma}, which is the theoretical curve for $\sg$ as a function of  $\imra$ for circular systems, derived in \citetalias{ORiordan2019}. The figure shows that any system in Regime 1 lies on this curve, when plotted using elliptical coordinates, regardless of the ellipticity of the lens. Although we are unable to derive a complete analytical treatment of two-image systems for elliptical lenses, this plot shows that the nature of the constraint on $\gamma$ for these systems is the same as for circular lenses, for which we do have a complete analytical treatment. This means that in Regime 1 the origin of the constraints on $\gamma$ is the complementarity of the constraints from the image positions and the constraints from the fluxes, breaking the degeneracy between $\gamma$ and source position $\beta$. While this is a satisfying explanation, there is in fact a subtlety hidden here, which we examine in \cref {sec:two-image-paradox}.

Consider now a row in \cref {fig:lens-grid}, which are mock observations of fixed ellipticity, with source position $\beta$ increasing from left to right. As $\beta$ increases, $\imra=\theta_{\rm min}/\theta_{\rm max}$ decreases, as is evident in the figure. 
In \cref {fig:full-results} systems with the same value of $\varepsilon$ are joined by a dotted line. Moving left to right along a row in  \cref {fig:lens-grid} corresponds to moving along a dotted line from right to left in  \cref {fig:full-results}. Sources at the right-hand end of a line of constant ellipticity in \cref {fig:full-results} are the aligned systems with $\beta=0$, i.e. four-image Einstein crosses. Following a single track of ellipticity in this figure, as $\beta$ increases, and $\imra$ decreases, we see that the constraint on $\gamma$ first improves ($\sg$ decreases), is best for a source near to but inside the caustic, and then worsens ($\sg$ increases) as the image multiplicity changes from four to two. The systems with the best constraints are four-image systems possessing two distinct images near the critical line, such as panels five and six in the top row of \cref {fig:lens-grid}. It is also evident from \cref {fig:full-results} that an aligned system, $\beta=0$, achieves a similar $\sg$ to a two-image system of the same ellipticity. 

The effect on these results of the size of the source is illustrated in \cref{fig:source-size}. For the circular case, \citetalias{ORiordan2019}, we showed that the constraint $\sg$ was independent of source size, provided the source size is substantially smaller than the Einstein angle. The figure plots $\sg$ against $\beta$, for a lens ellipticity $\varepsilon=0.5$, for three different source sizes $\reff=0.05, 0.1, 0.2 \arcsec$ (recall that the points in the main plot, \cref {fig:lens-grid}, are for $\reff=0.1 \arcsec$). The points for $\reff=0.05 \arcsec$ may not be as accurate, because the cuspy S\'{e}rsic profiles are not  as well sampled by the 0.04 arcsec pixels. It can be seen that size has no effect on $\sg$ at large values of $\beta$, as would be expected since these systems conform to the theory for the circular case.  However in the region of the best constraints, where the curves dip down, size has a significant effect. These are four-image systems where the source is near to but inside the caustic. In this region the constraints $\sg$ improve significantly for smaller source sizes.  At $\beta=0$ there is at most only a small effect of size.

\begin{figure*}
	\includegraphics[width=1.0\textwidth]{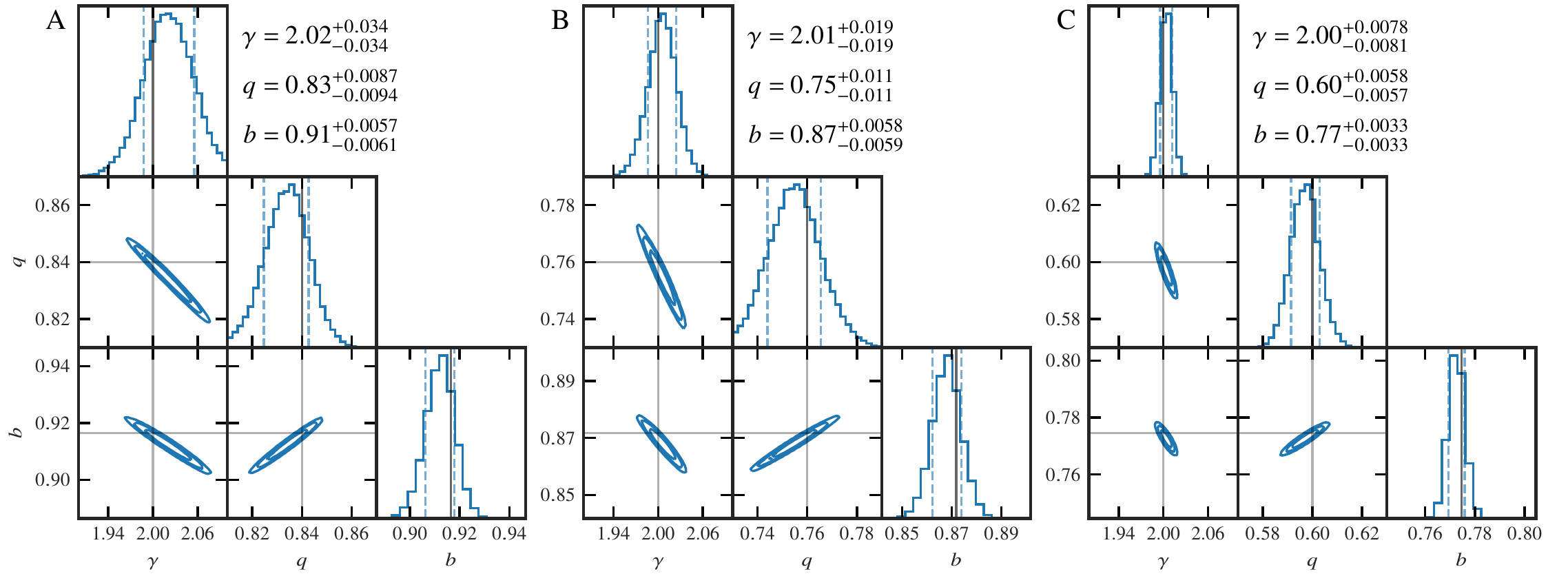}
	\caption{\label{fig:corner-plots} Examples of the posterior densities in $\gamma$, $q$ and $b$ for a system from each of the three regimes we have identified. System A is a two-image system and belongs to regime 1. System B has the source on the axis and so belongs to regime 2. System C is a four-image system with the source just inside the caustic and belongs to regime 3. In the 1D marginal posteriors the dashed lines are the 16th and 84th percentiles. In the 2D posteriors the contours indicate the 68\%, 95\% and 99\% credible regions. To aid comparison, all plots use the same size of range in parameter values, centred on the input values which are indicated by the solid lines. These same three systems are highlighted in \cref{fig:lens-grid}.}
\end{figure*}

\section{Regime 2: Aligned Four-Image Systems}
\label{sec:aligned-systems}

To understand the behaviour of four-image systems in \cref  {fig:full-results}, we begin with aligned systems, Regime 2, which are tractable analytically. We can derive an expression for the uncertainty on the slope in aligned systems by calculating the image positions and fluxes when $\beta=0$. Our treatment of these Einstein cross systems follows a similar route to our treatment of circular lenses in \citetalias{ORiordan2019}. Although these systems nominally have 12 observables (eight positions, four fluxes), because the system is symmetric there are in fact only four relevant observables: the image angular positions along the principal axes, $\theta_1$ and $\theta_2$, and their fluxes $f_1$ and $f_2$. With two angles and two fluxes, we are in the same situation as we were in analysing a circular lens. These four observables are sufficient to determine the four relevant parameters; $b,\gamma,q$ and $f_{\mathrm s}$.

\begin{figure*}
	\includegraphics[width=1.0\textwidth]{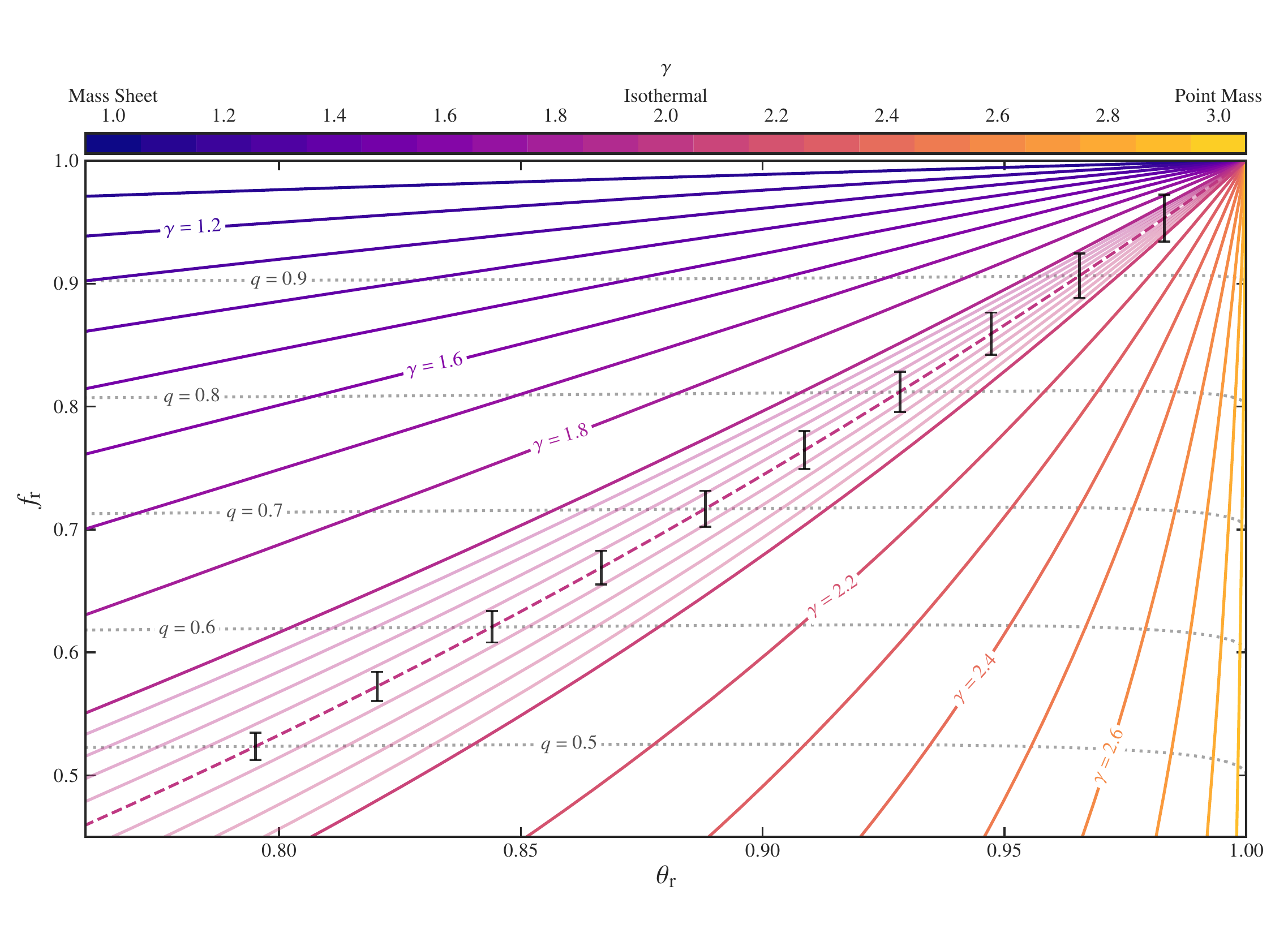}
	\caption{\label{fig:r2-obs-gamma-contours} Contours of $\gamma$ in the space of the observables $\imra$ and $\flra$ for a system with the source fixed to the origin. The slope is plotted in increments of $\Delta\gamma=0.1$ generally and in increments of $\Delta\gamma=0.02$ for $1.9<\gamma<2.1$. The dashed line is the isothermal ($\gamma=2$) contour. Error bars for the flux ratio are given by \cref  {eq:flux-ratio-error} (see Section 2.4 of \citetalias{ORiordan2019}). The dotted lines are contours of constant axis ratio with the $q=1$ contour coincident with the line $\flra=1$.}
\end{figure*}
With the source on the origin the images are fixed to the principal axes, so the inner image (at $\varphi=0$) has $z_1=\theta_1$ and the outer image (at $\varphi=\pi/2$) has $z_2=\iu\theta_2$.
There is only one component of the deflection angle at each image so we have
\begin{align}
	\alpha(z_1)&=\alpha_x(z_1)=\frac{2b}{1+q}\left(\frac{b}{q\theta_1}\right)^{\gamma-2}f(\gamma,q),\\
	\alpha(z_2)&=\iu\alpha_y(z_2)=\frac{2b}{1+q}\left(\frac{b}{\theta_2}\right)^{\gamma-2}f(\gamma,1/q),
\end{align}
where
\begin{equation}
	f(\gamma,q)={}_2F_1\left(1,{\textstyle \frac{\gamma-1}{2}};{\textstyle \frac{5-\gamma}{2}};{\textstyle -\frac{1-q}{1+q}}\right).
\end{equation}
The lens equation, \cref  {eq:lens-equation}, gives the image positions when $z'=0$ as a function of $b$, $\gamma$ and $q$,
\begin{align}
	\label{eq:image-positions-1}
	\theta_1&=b\left[\frac{2f(\gamma,q)}{(1+q)q^{\gamma-2}}\right]^{1/(\gamma-1)},\\
	\label{eq:image-positions-2}
	\theta_2&=b\left[\frac{2f(\gamma,q^{-1})}{1+q}\right]^{1/(\gamma-1)}.
\end{align}
We can eliminate $b$ by using the position ratio $\imra=\theta_1/\theta_2$,
\begin{equation}
	\label{eq:position-ratio}
	\imra^{\gamma-1}=\frac{f(\gamma,q)}{f(\gamma,q^{-1})}q^{2-\gamma}.
\end{equation}
The flux at each image is
\begin{align}
	f_1=f_\mathrm{s}|\mu(z_1)|&=f_\mathrm{s}\left|(\gamma-3)(\gamma-1)\left[(b/q\theta_1)^{\gamma-1}-1\right]\right|^{-1},\\
	f_2=f_\mathrm{s}|\mu(z_2)|&=f_\mathrm{s}\left|(\gamma-3)(\gamma-1)\left[(b/\theta_2)^{\gamma-1}-1\right]\right|^{-1}.
\end{align}
Dividing one by the other and substituting \cref  {eq:image-positions-1,eq:image-positions-2} for the image positions gives the flux ratio $\flra=f_1/f_2$ as
\begin{equation}	
	\label{eq:flux-ratio}
	\flra = \frac{f(\gamma,q)}{f(\gamma,q^{-1})}\frac{\left[2f(\gamma,q^{-1})-(q+1)\right]}{\Big[(q+1)/q-2f(\gamma,q)\Big]}.
\end{equation}
\Cref {eq:position-ratio,eq:flux-ratio} provide the relations between the observables $\imra$ and $\flra$, and the two relevant parameters $\gamma$ and $q$. By numerically inverting these equations we can plot contours of $\gamma$ and $q$ in the $(\imra,\flra)$ space. The results are provided in \cref  {fig:r2-obs-gamma-contours}, illustrating that from measurements of $\imra$ and $\flra$ for an Einstein cross system, the two parameters $\gamma$ and $q$ are uniquely determined. The procedure we have followed here is similar to the procedure used in \citetalias{ORiordan2019} to determine $\gamma$ and $\beta^\prime=\beta/\theta_2$. Equations (12) and (17) in the earlier paper are the analogues of \cref  {eq:position-ratio,eq:flux-ratio} here.

As we did for the circular case, we can use \cref  {eq:position-ratio,eq:flux-ratio} to find the uncertainty on the slope. We start with the assumption that the positions are measured to a much higher precision than the fluxes. In \citetalias{ORiordan2019}'s Section 2.5 we showed that the ratio between the position ratio and flux ratio uncertainties is $\sim \reff/\erad$. For the mock observations we use here this ratio is $\sim 0.1$ and so it is safe to assume that, compared to the flux ratio, the position ratio is essentially fixed when constraining the slope. This is equivalent to moving along a vertical line in \cref  {fig:r2-obs-gamma-contours}. 

Under this assumption the uncertainty on the slope is 
\begin{equation}
\sg=\left|\dv{\gamma}{\flra}\right|\sigma_\flra,
\end{equation}
where
\begin{equation}
\label{eq:flux-ratio-error}
\sigma_\flra=\frac{1}{S}(\flra + 1)\sqrt{\flra}.
\end{equation}
Because $S$ is the total S/N for all four images, this accounts for the fact that there are two independent measurements of each quantity $\theta_1, \theta_2, f_1, f_2$.

By considering a curve of constant $\imra$ in the $(\gamma,q)$ plane, it can be shown that
\begin{equation}
\dv{\flra}{\gamma} = \pdv{\flra}{\gamma} - \pdv{\flra}{q} \pdv{\imra}{\gamma}/\pdv{\imra}{q},
\end{equation}
and similarly for the axis ratio
\begin{equation}
\dv{\flra}{q} = \pdv{\flra}{q} - \pdv{\flra}{\gamma} \pdv{\imra}{q}/\pdv{\imra}{\gamma}.
\end{equation}
The partial derivatives are obtained by differentiating \cref  {eq:position-ratio,eq:flux-ratio}. The final expression for $\sg$ is too complex to be included here but its value for any pair $\imra, \gamma$ can be easily computed. In this way the curve of $\sg$ as a function of $\imra$ may be derived for any value of $\gamma$.

\begin{figure*}
	\includegraphics[width=1.0\textwidth]{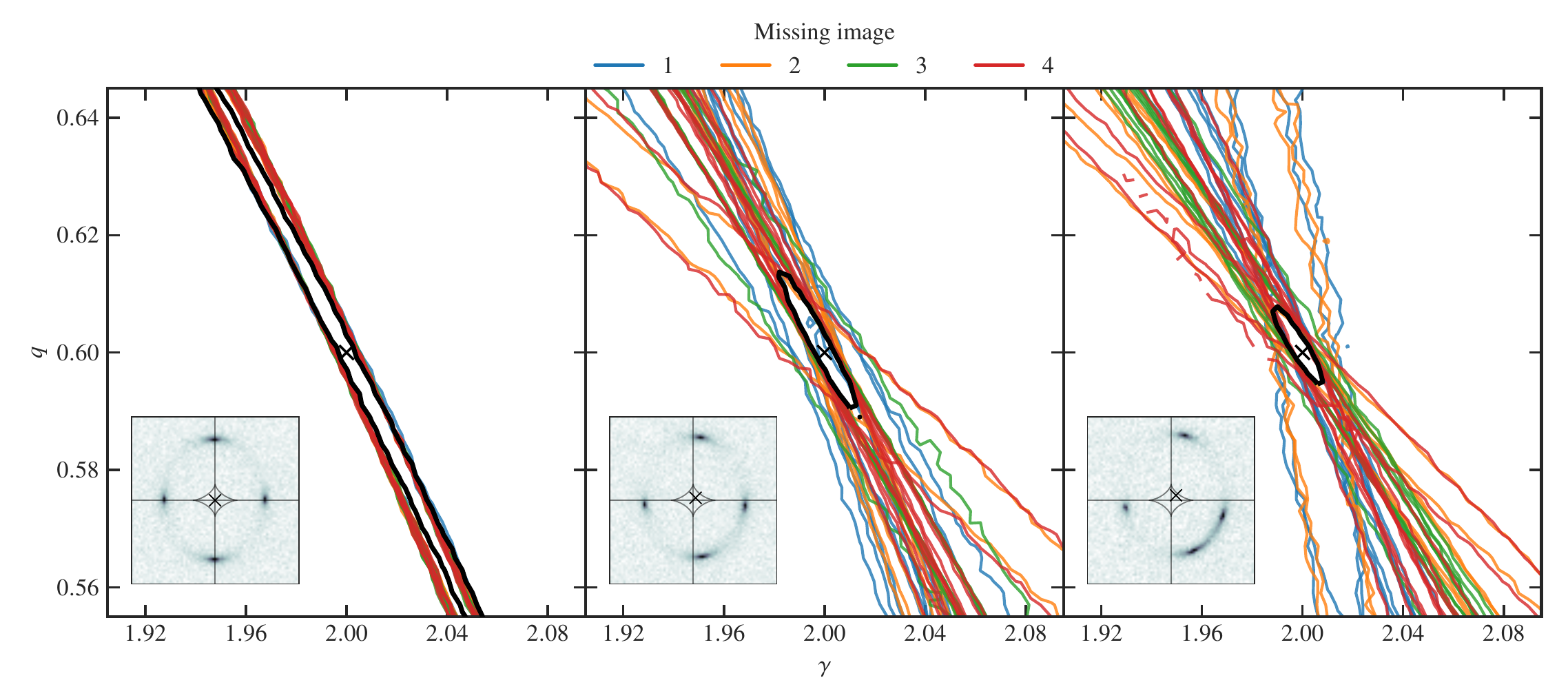}
	\caption{\label{fig:r3-psm-constraint} The complementary constraints from image positions, excluding fluxes, for three four-image systems with {\em (left, centre, right)} $\beta=(0.0,0.06,0.12)\arcsec$, each with $q=0.6$. The coloured contours are the $68\%$ posterior density confidence regions using different combinations of five of the eight image $x,y$ positions, as described in the text. The black contours are the results for all eight image $x,y$ positions. The insets show the equivalent extended-source mock observation for the same parameters.}
\end{figure*}

The result for $\gamma=2$, the value chosen for the mock observations, is plotted as the black dashed curve in \cref  {fig:full-results}. We expect this to match the mock observations for the case $\beta=0,$ i.e. at the RH end of each curve of constant $\varepsilon$. It can be seen that there is a good match between theory and simulation for larger values of ellipticity $0.3<\varepsilon<0.5$, but as $\varepsilon$ decreases the results for the mock observations lie systematically below the line i.e. the constraints are better than the theory predicts.
Referring to \cref  {fig:sple-diagram}, and considering $\beta=0$ i.e. the red images, one observes that for fixed source size as $\varepsilon$ decreases, the images become increasingly extended and contain higher-order information in addition to position and flux. It seems likely that it is this extra information which explains why the mock observations have smaller uncertainties, and lie below the theoretical curve. The same phenomenon was noted in discussing \citetalias{ORiordan2019}'s Figure 4.

This theory of aligned systems, $\beta=0$, predicts that $\sg$ is independent of source size, provided $\reff\ll\erad.$ As noted in the previous section, with reference to \cref{fig:source-size}, in the mock observations there is at most only a small effect of size at $\beta=0$, and the results may be consistent with no effect since, looking at the curve at large values of $\beta,$ there is a hint that the blue points are systematically slightly low due to sampling effects. 

\section{Regime 3: The general four-image system}
\label{sec:point-source-model}
While the behaviour in Regimes 1 and 2 may be understood from theoretical considerations, the behaviour of the general four-image system, Regime 3, is more complicated and is not amenable to an analytical treatment. In Regimes 1 and 2 the constraints come from combining all the position and flux information. For these two regimes the results are independent of the structure of the source, since the theory, which makes no assumptions about the structure of the source, matches the mock observations well. The situation in Regime 3 is different because the problem is effectively over-constrained. This may be understood by considering a compact source producing four images. Suppose one chooses to fit to the measured positions and fluxes, rather than to the full pixel information. Then the only relevant information about the source is the position $(\beta_x, \beta_y)$ and the flux $f_\mathrm{s}$. The lens is parameterised by $b, \gamma, q, \posa$. Each image offers a triplet of observables, two positions $(\theta_{x_i}, \theta_{y_i})$ and a flux $f_i$. There are therefore twelve constraints for seven parameters, leading to the possibility of poor fits and meaningless parameter estimates if the real lens cannot be well described by the elliptical mass models.

To understand how these constraints combine in the measurement of $\gamma$ we have developed a simplified treatment of the general problem. First we use the SPLE model (\cref{sec:theory}) to compute positions and fluxes analytically for any particular combination of the seven parameters. Next we create synthetic observables by adding appropriate uncertainties. We then fit to the observables using the same model, and compute the uncertainty $\sg$. We call this the  `position/flux model' to distinguish the method from mock observations where we fit to the pixel data. The great advantage of this treatment is that we can select which observables to use in making the fit. For example we could ignore all the flux information, or we could remove the position information for one or more of the images. This allows us to determine exactly which observables contribute to constraining the slope. We can also use this model to understand the transition between Regime 3 and Regime 2 and between Regime 3 and Regime 1. The results reveal that the three regimes each differ in terms of the information that constrains $\gamma,$ whether positions, fluxes, or combinations of the two, and whether or not the source is extended.

\subsection{Position/flux model}

We now briefly describe the process for creating a position/flux observation and constraining its parameters. From the above, the model has seven parameters, ${\bm p} = (\beta_x, \beta_y, f_\mathrm{s}, b, \gamma, q, \posa)$. For a given set of lens and source parameters we compute the image positions $(\theta_{x_i},\theta_{y_i})$ by numerically finding the two or four values of $z$ for which
\begin{equation}
	|z'+z-\alpha(z)|=0.
\end{equation}
The image flux $f_i$ is then found by multiplying the source flux $f_\mathrm{s}$ by the magnification at each image, i.e., $f_i=f_\mathrm{s}|\mu(\theta_{x_i},\theta_{y_i})|$, given by \cref  {eq:magnification}. We then add an amount of noise $n$ to each observable, where $n$ is drawn from a normal distribution with a variance determined as appropriate for the observable. The variances for the positions and fluxes are based on the noise model described in Section~2.2 of \citetalias{ORiordan2019}. The noise model makes the assumption that the variance in any image is proportional to the size, and therefore proportional to the flux. i.e. $\sigma_f^2=af$. The total signal to noise ratio is then
\begin{equation}
	S=\frac{\sum_{i}^{\nim}f_i}{\sqrt{\sum_{i}^{\nim}af_i}},
\end{equation}
where $\nim$ is the number of images. We fix $S$, allowing us to eliminate $a$ and find
\begin{equation}
	\label{eq:flux-error}
	\sigma^2_{f_i} = \frac{f_i}{S^2}\sum_{j}^{\nim}f_j.
\end{equation}
The positional uncertainties are more complicated since they depend on the shape and orientation of the magnified images. The purpose of applying the position/flux model is to understand where the constraints come from, rather than to reproduce accurately the uncertainties found in the mock observations. For this reason we use a simple prescription for the position uncertainty.
For the circular case, and for $\gamma=2$, the uncertainty on the radial position of the image is
\begin{equation}
	\label{eq:position-error}
	\sigma_{\theta_i}= \frac{\reff}{S_i}.
\end{equation}
This is the characteristic uncertainty on the image centroid, in any direction. For simplicity we assume that the positional uncertainties on each axis are given by this equation. With this assumption we ensure that the uncertainties on position ratios are much smaller than the uncertainties on flux ratios (by $\sim\reff/\erad$, see \citetalias{ORiordan2019} for details).
Our data is then a set of $3\nim$ `observations', two positions and one flux for each image, to which appropriate noise has been added.

We then constrain the model parameters following the method in Section 3 of \citetalias{ORiordan2019}, with one important difference. The likelihood, $\propto  \exp(-\chi^2/2),$ no longer compares measured and predicted pixel values but `observables': image positions and fluxes. The data now take the form of estimates of these, which for the $i$'th image are the measured position, $(\hat{\theta}_{x_i},\hat{\theta}_{y_i})$, and the measured flux, $\hat{f}_i$. The model, specified by the seven parameters ${\bm p}$, gives predictions for these observables, $\theta_{x_i}({\bm p})$, $\theta_{y_i}({\bm p})$ and $f_i({\bm p})$. Then $\chi^2$ is given by
\begin{equation}
	\chi^2 = \sum_{i}^{N_\mathrm{im}}\left\{\frac{[\hat{\theta}_{x_i}-\theta_{x_i}({\bm p})]^2 + [\hat{\theta}_{y_i}-\theta_{y_i}({\bm p})]^2}{\sigma_{\theta_i}^2} + \frac{[\hat{f}_i-f_i({\bm p})]^2}{\sigma_{f_i}^2}\right\}.
\end{equation}
Any of the observables for any of the images may be removed from the fit simply by excluding the relevant term from the $\chi^2$ sum.

\subsection{General four-image systems}
\label{sec:general-systems}

Following each track of ellipticity in \cref  {fig:full-results} shows that the best constraints on the slope occur if the source is inside and close to the caustic. These systems have the most highly magnified images in the population but, as we will show in this section, the strong constraints do not come from the flux information. Rather, these systems' image positions each provide complementary constraints on the slope which, when combined, are powerful enough to constrain the slope without flux information.

We show this using the position/flux model. In a general four image system, where $\beta\neq0$, there are 12 observables and seven parameters. If we discard the fluxes, in principle we still have enough observables (eight) to constrain the model. In such a case, one might expect the loss of flux information to negatively impact $\sg$. In fact, the loss of flux information does not significantly change $\sg$ and we find that the positional information is solely responsible for constraining the slope.

The reason for this becomes clear when considering the constraints provided by individual images, something we can do explicitly using the position/flux model. If we consider positions only then there are six parameters (dropping the source flux $f_\mathrm{s}$), and eight observables. By discarding three of the positions and fitting a six parameter model to a system with five observables, we force the solution into a degeneracy between the parameters. If we choose to drop one image's pair of positions, plus one position from another image, there are $4\times6=24$ possible degenerate fits of this kind for a four image system. 

\Cref {fig:r3-psm-constraint} shows the results of this exercise for three systems; one with the source aligned with the axis, one with the source just off the axis, and one with the source approaching the caustic. In each panel the degeneracy between $q$ and $\gamma$ is shown for the 24 combinations of position information. The colours correspond to which image is fully removed i.e. in each panel there are six degeneracies of the same colour. The black contour is the error ellipse when all eight positions are used. 
For the centre and right panels, where the source is off axis but inside the caustic, the plots make clear how the complementary information from the positions of the different images combine to provide the overall constraint on the parameters. Adding the flux information does not change the final black contours, showing that the constraints come from position information alone, because positions are measured more accurately than fluxes. Near the caustic the complementary constraints are less aligned and so the overall constraint is better. As the source moves closer to the origin, the different constraints become more closely aligned, and the overall constraint is worse.

\begin{figure}
	\includegraphics[width=1.0\columnwidth]{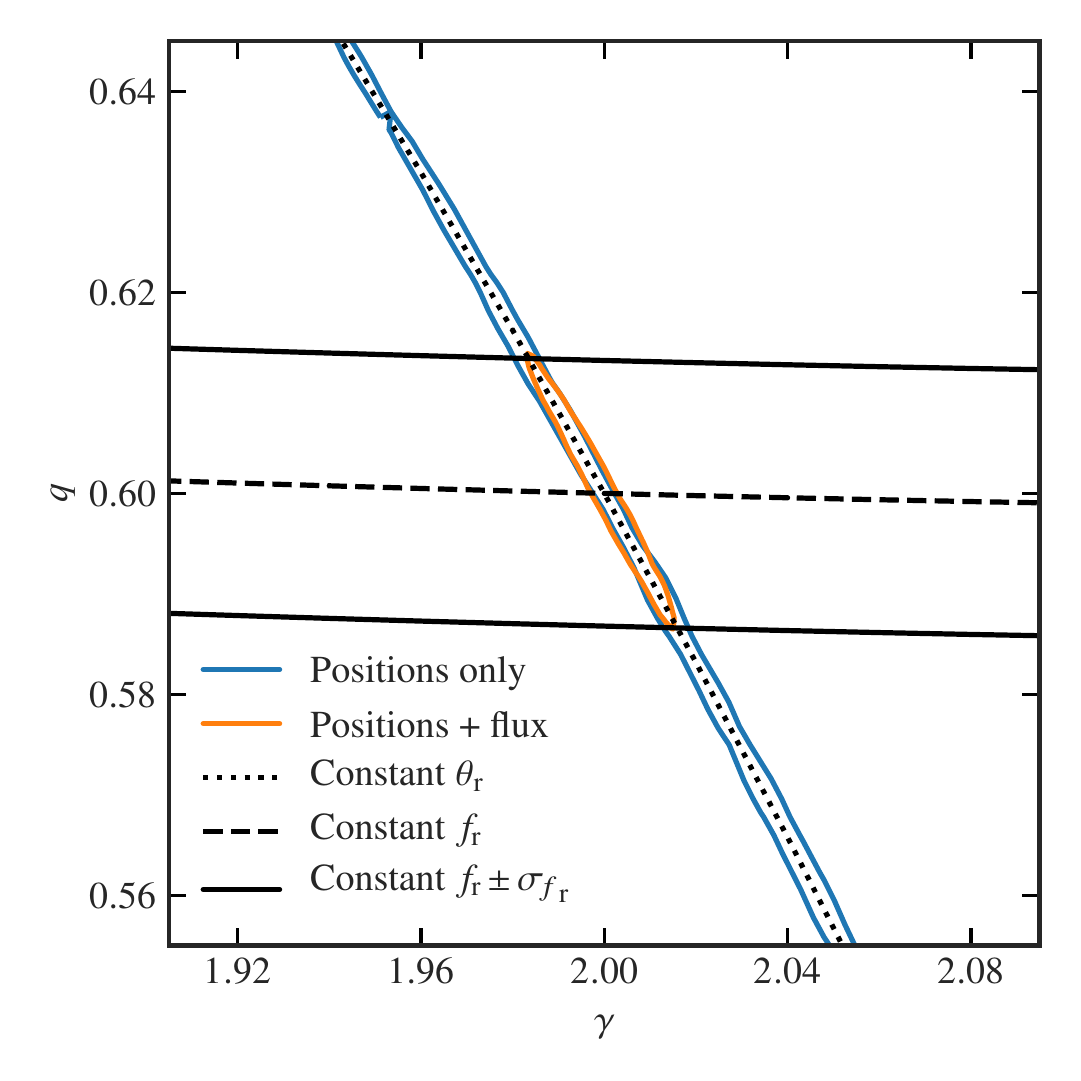}
	\caption{\label{fig:r2-flux-constraints} The $68\%$ posterior density credible regions for an aligned (Regime 2) system with $\varepsilon=0.4$. This system is identical to that in the leftmost panel of \cref{fig:r3-psm-constraint}. The blue contour shows the result from the position/flux model using positional information only, while the orange contour is the constraint when the flux constraints are added. The observables $\imra$ and $\flra$ are plotted as dotted and dashed line respectively.}
\end{figure}

The aligned system behaves as we would expect from the analysis in \cref  {sec:aligned-systems}. We showed previously that the positional information alone is not enough to constrain the slope and the figure explains why this is the case. Now all the 24 degeneracies line up to the same degenerate curve and so adding all the positional information together is still not enough to constrain $\gamma$ and $q$, despite there being more observables than parameters. This is made explicit by the black contour which shows the constraints when using all eight positions. The degeneracy is only broken by adding the flux information.

The importance of flux information for Regime 2 is made clear by reference to \cref{fig:r2-flux-constraints}.  Here the blue contour is the degeneracy between $\gamma$ and $q$ found using the position/flux model with the eight positional constraints only (this is the same as the black contour in the left panel of 
\cref{fig:r3-psm-constraint}). The orange ellipse is the constraint when the flux information is added in. In this particular figure the only model parameters are $\gamma$ and $q$. Therefore we can calculate the observables $\imra$ and $\flra$ analytically everywhere in the plane using \cref{eq:position-ratio,eq:flux-ratio}. Contours of these observables make the origin of the constraint clear. The dotted line is the contour of constant $\imra$, which matches the degeneracy. The black dashed line is the line of constant $\flra$ corresponding to this configuration, and the solid black lines show the $1\sigma$ uncertainties, which agree perfectly with the position/flux model result. This plot shows why the final uncertainty $\sg$ is controlled by the uncertainty on the flux ratio.

These results, together, provide an explanation for the behaviour shown in \cref{fig:source-size}. There we found that there was little if any effect of source size on
 $\sg$ in Regimes 1 and 2. Using the position/flux model we have shown that in Regime 3 the constraints on the value of $\gamma$ come from positional information alone. Since the uncertainty on position scales as $\reff$ (\cref{eq:position-error}), this predicts that in Regime 3 we expect the approximate relation $\sg\propto\reff$. Referring to  \cref{fig:source-size} we see that near the minimum this is indeed approximately the case in that at the minimum $\sg$ for $\reff=0.05\arcsec$ is nearly four times smaller than for $\reff=0.2\arcsec$. These results show how in Regime 3 the uncertainty on $\gamma$ is controlled by how accurately positions are measured, and that there is a transition in moving to Regimes 1 and 2 where the uncertainty on $\gamma$ is set by how accurately fluxes are measured.

\subsection{Two image systems}
\label{sec:two-image-paradox}

\begin{figure}
	\includegraphics[width=1.0\columnwidth]{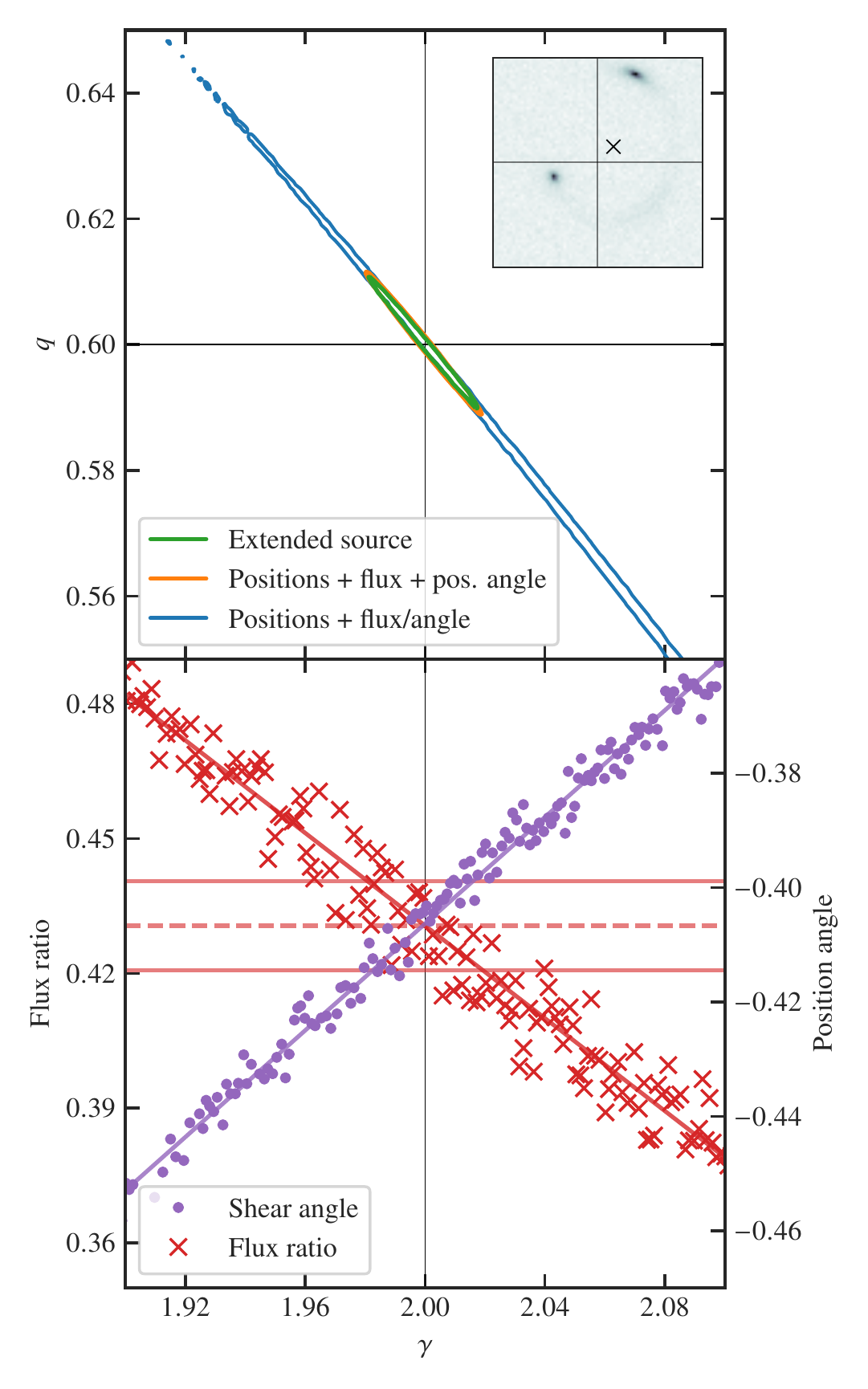}
	\caption{\label{fig:r1-psm-constraint} {\em Upper:} The constraint on the slope in the Regime 1 system shown in the inset. This system has $q=0.6$ and $\beta=0.16$ arcsec. The upper frame shows the $68\%$ posterior density credible regions for the three fits listed in the key. {\em Lower:}  Circles and crosses are representative values of the respective missing observable along the blue contour i.e. position angle for the positions+flux fit, and flux ratio for the positions+angle fit. The red horizontal lines mark the 1$\sigma$ uncertainty on the flux ratio, given by \cref{eq:flux-ratio-error}. 
	}
\end{figure}

We noted in \cref{sec:introduction} that the two-image systems at any ellipticity (Regime 1) agree with the theory we established in \citetalias{ORiordan2019} for circular systems. For circular systems it is the combination of positional and flux information which provides a constraint on the slope. Nevertheless the uncertainty on the constraint $\sg$ ultimately depends only on the uncertainty on the flux measurement. This is because the positions are measured to a much higher precision relative to the fluxes and they can be treated as fixed quantities when determining the slope. Evidently the same is true for  two-image systems in the elliptical case. Yet further consideration of the position/flux analysis reveals an apparent paradox. A generic two-image elliptical system in Regime 1 has seven parameters $(\beta_x, \beta_y, \srcf, b, \gamma, q, \posa)$ but only six observables (four positions, two fluxes). Running the position/flux model for a configuration in Regime 1 indeed shows the system constrained only to a degenerate line in parameter-space. An example of this is shown by the blue contour in \cref{fig:r1-psm-constraint}. 

It is evident that the disagreement between the results from the mock observations and those from the position/flux analysis comes about because the source is extended in the mock observations. Since the value of $\sg$ is controlled by the flux uncertainties, we can reconcile these results if we postulate that in the mock observations there is additional information of a positional nature contributing to the fit. We can show this for the simulations used here if we identify the position angle of the outer image, i.e. the direction of the shear, as an additional observable (the information from the outer image will dominate over that from the inner image). It is interesting to note that image position angle is one of the observables used in the model-independent approach of \citet{Wagner1}, described in the introduction. Since this is a positional quantity we will assume that the uncertainty is small.
The position angle is readily calculated from the expression for the complex shear. If $\eta(z)=\eta_1 + \iu \eta_2$ is the complex shear then the position angle $\varphi$ of a small image at $z$, from a circular source, is given by the solution of the two equations
\begin{eqnarray}
	\eta_1&=&c\,\cos[2\varphi(z)], \\
	\eta_2&=&c\,\sin[2\varphi(z)],
\end{eqnarray}
where $c$ is a positive constant.

 The result of adding the image position angle to the fit is illustrated in \Cref{fig:r1-psm-constraint}. As noted above, fitting the seven parameter position/flux model to the six original observables (four positions and two fluxes) produces the blue contour, a degenerate curve with no constraint on the slope. If instead we use only the positional information, the four positions and the position angle, we now have five observables and six parameters (we have discarded the source flux). Fitting with this combination produces an identical degeneracy to the blue contour. Along the blue contour, i.e. the degeneracy between $\gamma$ and $q$ for both a positions+flux and positions+angle fit, we take representative points from the MCMC samples and select the missing observable not used in the fit, either the position angle or the flux ratio. In the lower frame we plot the respective missing observables at the sample value of $\gamma$, showing that each quantity varies along the degeneracy.

Combining all seven observables (four positions, two fluxes, one angle) breaks the degeneracy and constrains the slope. This is plotted as the orange contour. The constraints match those from the extended-source mock observation, plotted as the green contour. For this calculation the uncertainty we used on the position angle was arbitrarily small (in line with it being a positional variable). This ensures that the flux measurement is still the dominant source of error, and in all cases the uncertainty on the flux is set by the fixed total S/N in the images. These results indicate that it is indeed the additional position information provided by the extended source that reconciles the two calculations. We hence recover the same behaviour as in the circular systems of \citetalias{ORiordan2019}, explaining the close match between the Regime 1 results and \cref{eq:sigmagamma} in \cref{fig:full-results}. Despite the larger number of model parameters and observables the constraint on the slope still comes from a combination of positional and flux information, with the precision of the constraint determined by the uncertainty on the flux.

\section{Discussion and summary}
\label{sec:conclusions}
In this paper we have extended the analysis of circular systems in \citetalias{ORiordan2019} to consider how strong gravitational lenses can be used to measure the slope $\gamma$ of a power-law density distribution in elliptical lenses, the SPLE model. We have found that the manner in which the observables constrain the measured slope $\gamma$ is different in three separate regimes.

Systems with two-images define Regime 1. In this regime the measured uncertainty on the slope $\sg$ was found to match the theoretical prediction for circular systems, developed in \citetalias{ORiordan2019}, when transformed to elliptical coordinates. Also, in contrast with the other two regimes, in Regime 1 it is only possible to measure the slope if the images are resolved, because otherwise there are insufficient constraints. In this regime the constraints rely on the combination of position and flux information, with the size of the slope uncertainty determined by the precision of the flux measurements. This explains why $\sg$ measured for these more complicated elliptical systems matches the prediction of the simple analysis for circular systems.

A four-image system in general gives much more accurate constraints than a two-image system with the same lens ellipticity, and the same total image S/N, by a factor of two to eight, depending on the source size. We split the four-image systems into two regimes. Regime 2 concerned the special case of a source fixed to the axis, producing an Einstein cross image. For these systems we derived analytical expressions for $\gamma,$ $q$ and $\sg$ as a function of the ratios between the minor and major axes of the positions and of the fluxes, and their uncertainties. In this regime the flux uncertainties again dominate the error budget.

Finally, we analysed the best-constrained systems, those of Regime 3. These are the systems with the source inside but close to the caustic, producing four bright images embedded in a ring. In these systems the measurements of the image positions alone are enough to constrain the slope because each image position offers complementary information on the mass profile. In Regime 2 the positional information alone provides only degenerate constraints, but as the source moves away from the axis the constraints from each image become successively more complementary, reaching a maximum just before the source crosses the caustic.

In these first two papers in the series we have investigated the general constraining power of strong lensing information for power-law mass profiles. In both papers we have identified some common results. We showed that for a fixed S/N, the relative uncertainties on the positions of the images are smaller than those of the fluxes, and they depend on the sizes of the images, which depend in turn on the size of the source and the resolution of the observations. Since flux errors are larger relatively, when flux information is important for constraining the slope, as in Regimes 1 and 2, it controls the value of $\sg$, and therefore the structure and size of the source (which determine the positional uncertainties) are largely irrelevant. In Regimes 1 and 2, in cases where the source structure is complex, the extra information does not provide additional constraints on the mass profile. Rather it can be used to produce a more complex model of the source. In Regime 3, where the positional information constrains the slope without flux, the constraint on the slope will depend on the size and structure of the source because a larger source makes the image positions less accurate (for fixed total S/N).  

Regime 3 is the most interesting for measuring the mass profile of a gravitational lens. For four-image systems with $S=100,$ we have shown that measurement of $\gamma$ to an accuracy of $\sg<0.01$ is feasible. Given this, for codes that combine full surface-brightness-fitting of gravitational lensing data with dynamical information, such as CAULDRON \citep{Barnabe2007,Barnabe2009}, if the images are of high S/N it would appear that the accuracy of the measurement of $\gamma$ will be strongly dominated by the lensing fit. 

Since in Regime 3 only the positional information contributes to the measurement of $\gamma$, this reveals the interesting possibility of constraining more complicated mass models because there are unused additional constraints provided by the four fluxes. In fact in principle resolved sources contain even more information. Positions and fluxes are determined by respectively the deflection angle and its first derivative, which correspond to the first and second derivatives of the potential. With resolved sources the radial curvature of the deflection angle, i.e. the third derivative of the potential, should also be measurable. This is in the same spirit as the model-independent approach \citep{Wagner1, Wagner2}, which uses multiple observables in addition to positions and fluxes. However in the model-fitting approach all the constraints are used simultaneously but implicitly through fitting the pixel data, rather than explicitly through measuring observables.

In Papers I and II we have obtained very good agreement between the results of analysing mock observations and the predictions of analytical treatments. This is valuable because in future papers we will rely on mock observations only, as the models become too complex for an analytical treatment.
In the next two papers in this series we will introduce and use a broken power-law model to determine the available constraints on the slope interior to the images and to measure the sensitivity of the images to the slope as a function of radius. 

\section*{Acknowledgements}
We thank the reviewer Nicolas Tessore for comments that improved the presentation of the paper. We are grateful to the Imperial College Research Computing Service for HPC resources and support. CMO'R is supported by an STFC Studentship.

\section*{Data Availability}
The data used in this paper are available from the corresponding author on request.

\bibliographystyle{mnras}
\bibliography{bibliography}

\bsp	
\label{lastpage}
\end{document}